\documentclass[a4paper,12pt]{article}
\usepackage{jheppubmod}
\usepackage{amsmath}
\usepackage{amssymb}
\usepackage{bbm}
\usepackage{amsfonts}
\usepackage{tipa}
\usepackage{mathrsfs}
\usepackage{mathtools}
\usepackage{graphicx}

\title{Gauge Invariant Descriptions of Gluon Polarizations}

\author{Zhi-Qiang Guo}
\author{and Iv\'{a}n Schmidt}
\emailAdd{zhiqiang.guo@usm.cl}
\emailAdd{ivan.schmidt@usm.cl}
\affiliation{Departamento de F\'{i}sica y Centro Cient\'{i}fico
Tecnol\'{o}gico de Valpara\'{i}so,\\ Universidad T\'{e}cnica Federico
Santa Mar\'{i}a,\\ Casilla 110-V, Valpara\'{i}so, Chile}

\abstract{We propose methods to construct gauge invariant decompositions of the nucleon spin, particularly gauge invariant descriptions of the gluon polarization. We show that gauge invariant decompositions of the nucleon spin can be derived naturally from the conserved current of a generalized Lorentz transformation by the Noether theorem. We propose a gauge invariant extension of the Chern-Simons current as a possible description of the gluon spin, and we also examine the problem of gauge dependence of this extended Chern-Simons current.}


\begin{document}
\maketitle

\section{Introduction}\label{sec:1}

Since the experimental discovery that the quark spin only contributes a small portion to the nucleon spin~\cite{Ashman:1987hv}, the nucleon spin problem has attracted tremendous experimental and theoretical efforts over the past around two decades~\cite{Aidala:2012mv}. Nevertheless, how to formulate complete and gauge invariant decompositions of the nucleon spin and particularly how to formulate gauge invariant descriptions of the gluon polarization remains an open and interesting theoretical problem so far.

Regarding the decomposition of the nucleon spin, three popular decompositions have been proposed based on different grounds. The decomposition of Jaffe and Manohar~\cite{Jaffe:1989jz} is the conventional canonical one derived from the Noether theorem. The decomposition of Ji~\cite{Ji:1996ek} has the gauge invariant formulation by means of the improved energy-momentum tensor of Belinfante~\cite{Belinfante:1940q}, but it does not provide a gauge invariant description of the gluon spin. By decomposing the gauge field into its physical part and its pure gauge part, Chen et al.~\cite{Chen:2008ag} proposed a novel gauge invariant decomposition and a gauge invariant description of the gluon spin is also obtained.  The similarities and differences among these three decompositions have been examined extensively from theoretical perspectives~\cite{Leader:2011za,Wakamatsu:2012ve,Lorce:2012rr}. One purpose of this paper is to show that these three decompositions can be derived naturally from the conserved Noether currents induced by a generalized Lorentz transformation, although they have apparently different formulations. These results shall be given in Section~\ref{sec:3}.

Regarding descriptions of the gluon spin, among these three decompositions, only the decomposition of Chen et al. can provide gauge invariant expressions by using nonlocal formulations of the pure gauge field. This gauge invariant description can be regarded as the gauge invariant extension of the conventional canonical one. In this paper, we propose that there may exist possible alternative descriptions of the gluon spin. For this purpose, we construct a gauge invariant extension of the Chern-Simons current by using the pure gauge field defined by Delbourgo and Thompson~\cite{Delbourgo:1986wz}. This extended Chern-Simons current has the nice feature that it satisfies the same equation as the conventional Chern-Simons current. We also verify its gauge independence in the Schwinger model~\cite{Manohar:1990eu}. The relevant discussions shall be given in section~\ref{sec:2} and~\ref{sec:4}. We give some conclusions in section~\ref{sec:5}. More discussions on the gauge invariant decompositions from the Noether theorem and the extended Chern-Simons current as an appropriate description of the gluon spin will be given in a subsequent paper~\cite{Guo:2013jia}.

\section{Gauge Invariant Extensions of the Chern-Simons Current}\label{sec:2}

 In this section, we discuss the gauge invariant extension of the Chern-Simons current as a possible candidate of descriptions of the gluon spin, which avoids the gauge-dependence problem of the conventional Chern-Simons current. The Chern-Simons current $K^{\mu}$
\begin{eqnarray}
\label{chern-simons-current}
K^{\mu}=\frac{1}{2}\varepsilon^{\mu\nu\alpha\beta}A^{a}_{\nu}(F^{a}_{\alpha\beta}-\frac{g}{3}f^{abc}A_{b\alpha}A_{c\beta}),
\end{eqnarray}
which satisfies the equation
\begin{eqnarray}
\label{chern-simons-current-equation}
\partial_{\mu}K^{\mu}=\frac{1}{2}F^{a}_{\mu\nu}\tilde{F}_{a}^{\mu\nu},\hspace{2mm}
\tilde{F}^{a}_{\mu\nu}=\frac{1}{2}\varepsilon_{\mu\nu\alpha\beta}F_{a}^{\alpha\beta},
\end{eqnarray}
could be an appropriate description of the gluon polarization, as has been proposed in~\cite{Altarelli:1988nr,Carlitz:1988ab,Efremov:1988zh}, partly because it is connected to the anomalous equation for the axial-vector current
\begin{eqnarray}
\label{anomaly-equation}
\partial_{\mu}(\bar{\psi}\gamma^{\mu}\gamma^{5}\psi)=\frac{\alpha_{s}}{4\pi}F^{a}_{\mu\nu}\tilde{F}_{a}^{\mu\nu}.
\end{eqnarray}
Nevertheless, an unpleasant character of the Chern-Simons current is that it is not invariant under large gauge transformations. Under a gauge transformation
\begin{eqnarray}
\label{chern-simons-current-gauge-field-trans}
A_{\mu}{\rightarrow}UA_{\mu}U^{-1}-\frac{i}{g}(\partial_{\mu}U)U^{-1},
\end{eqnarray}
the Chern-Simons current transforms as
\begin{eqnarray}
\label{chern-simons-current-gauge-trans}
K_{\mu}{\rightarrow}K_{\mu}-\frac{2i}{g}\varepsilon_{\mu\nu\alpha\beta}\partial^{\alpha}\mathrm{Tr}(U^{-1}\partial^{\nu}UA^{\beta})
-\frac{2}{3g^2}\varepsilon_{\mu\nu\alpha\beta}\mathrm{Tr}\{U^{-1}(\partial^{\nu}U)U^{-1}(\partial^{\alpha}U)U^{-1}(\partial^{\beta}U)\},
\end{eqnarray}
which is not invariant if the gauge transformation is large. Therefore, calculations with the Chern-Simons current are surrounded by the problem of gauge dependence~\cite{Manohar:1990eu,Balitsky:1991te,Shore:1999be}. However, we can construct a gauge invariant extension of the Chern-Simons current to cure the gauge-dependence problem. We have assumed in our above discussion that the expression~(\ref{chern-simons-current}) is the unique solution of Eq.~(\ref{chern-simons-current-equation}). Now we could put forward the following question: Does there exist another expression of  $K^{\mu}$, which satisfies Eq.~(\ref{chern-simons-current-equation})? The answer is positive, as can be seen from the Abelian case. For a Abelian $U(1)$ theory, the Chern-Simons current is
\begin{eqnarray}
\label{chern-simons-current-abelian}
K^{\mu}=\frac{1}{2}\varepsilon^{\mu\nu\alpha\beta}A_{\nu}F_{\alpha\beta},
\end{eqnarray}
which satisfies
\begin{eqnarray}
\label{chern-simons-current-equation-abelian}
\partial_{\mu}K^{\mu}=\frac{1}{2}F_{\mu\nu}\tilde{F}^{\mu\nu},
\hspace{2mm}\tilde{F}_{\mu\nu}=\frac{1}{2}\varepsilon_{\mu\nu\alpha\beta}F^{\alpha\beta}.
\end{eqnarray}
But Eq.~(\ref{chern-simons-current-abelian}) is not the unique solution of Eq.~(\ref{chern-simons-current-equation-abelian}). We can construct another expression as follows{\footnote{We noticed that X.-S. Chen has proposed such kind of constructions in a workshop talk~\cite{Chen:2009mt}.}}
\begin{eqnarray}
\label{chern-simons-current-abelian-gie}
\mathcal {K}^{\mu}=\frac{1}{2}\varepsilon^{\mu\nu\alpha\beta}F_{\alpha\beta}(A_{\nu}-\partial_{\nu}\phi).
\end{eqnarray}
Obviously, the expression~(\ref{chern-simons-current-abelian-gie}) satisfies
\begin{eqnarray}
\label{chern-simons-current-equation-abelian-gie}
\partial_{\mu}\mathcal {K}^{\mu}=\frac{1}{2}F_{\mu\nu}\tilde{F}^{\mu\nu},
\end{eqnarray}
where the Bianchi identities has been used. It has the remarkable property that it is invariant under the gauge transformation
\begin{eqnarray}
\label{chern-simons-current-gauge-field-trans-ableian}
A_{\mu}{\rightarrow}A_{\mu}+\partial_{\mu}\Lambda,\hspace{2mm}\phi{\rightarrow}\phi+\Lambda.
\end{eqnarray}
The expression~(\ref{chern-simons-current-abelian-gie}) can be thought as a gauge invariant extension of the Chern-Simons current~(\ref{chern-simons-current-abelian}). Using the notation of Chen et al.~\cite{Chen:2008ag}, we can make the identifications
\begin{eqnarray}
\label{chern-simons-current-gauge-field-ableian-pure}
A^{\mu}_{\mathrm{pure}}=\partial^{\mu}\phi,\hspace{2mm}A^{\mu}_{\mathrm{phys}}=A^{\mu}-A^{\mu}_{\mathrm{pure}}.
\end{eqnarray}
The foregoing constructions can be generalized to the non-Abelian case. We propose the following expressions for the non-Abelian case
\begin{eqnarray}
\label{chern-simons-current-gie}
\mathcal {K}_{\mu}=\frac{1}{2}
\varepsilon_{\mu\nu\alpha\beta}A^{a\nu}_{\mathrm{phys}}(F^{a\alpha\beta}-\frac{g}{3}f^{abc}A^{b\alpha}_{\mathrm{phys}}A^{c\beta}_{\mathrm{phys}}),
\end{eqnarray}
where $F_{\mu\nu}=\partial_{\mu}A_{\nu}-\partial_{\nu}A_{\mu}-ig[A_{\mu},A_{\nu}]$ still remains the conventional expression, and
\begin{eqnarray}
\label{chern-simons-current-gauge-field-pure}
A^{\mu}_{\mathrm{pure}}=\frac{i}{g}V^{-1}\partial^{\mu}V,\hspace{2mm}A^{\mu}_{\mathrm{phys}}=A^{\mu}-A^{\mu}_{\mathrm{pure}}.
\end{eqnarray}
The expression~(\ref{chern-simons-current-gie}) is invariant under the gauge transformation
\begin{eqnarray}
\label{chern-simons-current-gauge-field-trans-gie}
A_{\mu}{\rightarrow}UA_{\mu}U^{-1}-\frac{i}{g}(\partial_{\mu}U)U^{-1}, \hspace{2mm}V{\rightarrow}VU^{-1}.
\end{eqnarray}
We can reformulate~(\ref{chern-simons-current-gie}) as
\begin{eqnarray}
\label{chern-simons-current-gie-reform}
\mathcal{K}_{\mu}&=&\varepsilon_{\mu\nu\alpha\beta}\mathrm{Tr}(F^{\alpha\beta}A^{\nu}_{\mathrm{phys}}
+\frac{2i}{3}gA^{\nu}_{\mathrm{phys}}A^{\alpha}_{\mathrm{phys}}A^{\beta}_{\mathrm{phys}})\\
\label{chern-simons-current-gie-reform-1}
&=&\varepsilon_{\mu\nu\alpha\beta}\mathrm{Tr}(F^{\alpha\beta}A^{\nu}+\frac{2i}{3}gA^{\nu}A^{\alpha}A^{\beta})
-\frac{2i}{3}g\mathrm{Tr}(A^{\nu}_{\mathrm{pure}}A^{\alpha}_{\mathrm{pure}}A^{\beta}_{\mathrm{pure}})\\
\label{chern-simons-current-gie-reform-2}
&-&2\varepsilon_{\mu\nu\alpha\beta}\mathrm{Tr}(\partial^{\alpha}A^{\beta}A^{\nu}_{\mathrm{pure}}
-igA^{\nu}A^{\alpha}_{\mathrm{pure}}A^{\beta}_{\mathrm{pure}}).
\end{eqnarray}
Employing the definition of  $A^{\mu}_{\mathrm{pure}}$ in Eq.~(\ref{chern-simons-current-gauge-field-pure}), we can show that the terms of Eq.~(\ref{chern-simons-current-gie-reform-2}) can be combined to be a total divergence, then we obtain
\begin{eqnarray}
\label{chern-simons-current-gie-reform-3}
\mathcal{K}_{\mu}&=&\varepsilon_{\mu\nu\alpha\beta}\mathrm{Tr}(F^{\alpha\beta}A^{\nu}+\frac{2i}{3}gA^{\nu}A^{\alpha}A^{\beta})
-\frac{2i}{g}\varepsilon_{\mu\nu\alpha\beta}\partial^{\alpha}\mathrm{Tr}[A^{\beta}V^{-1}\partial^{\nu}V]\\
&-&\frac{2}{3g^2}\varepsilon_{\mu\nu\alpha\beta}\mathrm{Tr}\{V^{-1}(\partial^{\nu}V)V^{-1}(\partial^{\alpha}V)V^{-1}(\partial^{\beta}V)\}.\nonumber
\end{eqnarray}
With Eq.~(\ref{chern-simons-current-gie-reform-3}), we can easily check that the expression~(\ref{chern-simons-current-gie}) also satisfies
\begin{eqnarray}
\label{chern-simons-current-equation-gie}
\partial_{\mu}\mathcal{K}^{\mu}=\frac{1}{2}F^{a}_{\mu\nu}\tilde{F}_{a}^{\mu\nu}.
\end{eqnarray}
Therefore, a gauge invariant extension of the non-Abelian Chern-Simons current can be constructed successfully. From Eq.~(\ref{chern-simons-current-gie-reform-3}), we see that the gauge invariant current includes the conventional Chern-Simons current and some terms contributed by the pure gauge field.  These contributions from the pure gauge field may be related to the gluon topology as considered by Bass~\cite{Bass:1997zz}. A problem to be solved is how to derive a manifest expression for the pure gauge field. This problem has been highlighted from a more physical perspective by Chen et al.~\cite{Chen:2008ag}, and several expressions of the pure gauge field have been proposed~\cite{Chen:2011gn,Hatta:2011zs,Zhang:2011rn,Lorce:2012ce}. However, we noticed that Delbourgo and Thompson~\cite{Delbourgo:1986wz} and Verschelde et al.~\cite{Verschelde:2001ia} have proposed manifest expressions for the pure gauge field through different grounds. In the proposal of Delbourgo and Thompson~\cite{Delbourgo:1986wz}, a pure gauge field can be obtained from the following equation
\begin{eqnarray}
\label{pure-gauge-laudau-pure}
D_{\mu}[A^{\mu}-\frac{i}{g}U^{-1}\partial^{\mu}U]=0,\hspace{2mm} U=\mathrm{exp}(-ig\varphi),
\end{eqnarray}
where $D_{\alpha}=\partial_{\alpha}-ig[A_{\alpha},\cdot~]$. This kind of equations can be solved by a formal series~\cite{Delbourgo:1986wz,Kubo:1986ps}. A manifest expression for $A^{\mu}_{\mathrm{pure}}$ can be given by
\begin{eqnarray}
\label{pure-gauge-laudau-phys-solution}
A^{\mu}_{\mathrm{pure}}=\partial^{\mu}\frac{1}{\partial^2}\partial^{\alpha}A_{\alpha}
+ig\partial^{\mu}\frac{1}{\partial^2}[A_{\beta},\partial^{\beta}\frac{1}{\partial^2}\partial^{\alpha}A_{\alpha}]+\mathcal{O}(g^2).
\end{eqnarray}
Employing this expression, we shall show in section \ref{sec:4} that the gauge invariant currents~(\ref{chern-simons-current-abelian-gie}) and~(\ref{chern-simons-current-gie}) can yield gauge-independent results.

\section{Gauge Invariant Decompositions from Noether Theorem}\label{sec:3}

In this section, we show that the decomposition of Jaffe and Manohar, the decomposition of Ji and the decomposition of Chen et al. can all be derived naturally from the Noether theorem, which reveals interesting relations among these three decompositions. We consider the Lagrangian of quantum chromodynamics~(QCD) with the $\theta$ term,
\begin{eqnarray}
\label{qcd-lag}
\mathcal {L}&=&-\frac{1}{4}F^{a}_{\mu\nu}F^{a\mu\nu}-\theta\frac{g^2}{32\pi^2}F^{a}_{\mu\nu}\tilde{F}^{a\mu\nu}\\
&+&\frac{i}{2}[\bar{\psi}\gamma^{\mu}(\partial_{\mu}-igA_{\mu})\psi-(\partial_{\mu}\bar{\psi}+ig\bar{\psi}A_{\mu})\gamma^{\mu}\psi],\nonumber
\end{eqnarray}
which is invariant under the gauge transformation
\begin{eqnarray}
\label{noether-gauge-trans}
\hat{A}_{\mu}=UA_{\mu}U^{-1}-\frac{i}{g}\partial_{\mu}UU^{-1},\hspace{2mm}\hat{\psi}=U\psi,
\end{eqnarray}
and the Lorentzian transformation
\begin{eqnarray}
\label{noether-lorentz-trans}
\hat{A}_{\mu}(x')=\Lambda_{\mu}^{\hspace{1mm}\nu}A_{\nu}(x),\hspace{2mm}\hat{\psi}(x')=S[\Lambda]\psi(x).
\end{eqnarray}
However, as considered by Bjorken and Drell~\cite{Bjorken:1965zz}, by Weinberg~\cite{Weinberg:1965rz,Weinberg:1995mt} and recently more thoughtfully by Lorc\'{e}~\cite{Lorce:2012rr}, the gauge field $A_{\nu}$ does not need to be transformed as a Lorentz vector, and the Lagrangian~(\ref{qcd-lag}) is actually invariant under a more general Lorentz transformation
\begin{eqnarray}
\label{noether-lorentz-gauge-trans-v}
\tilde{A}_{\mu}(\tilde{x})&=&\Lambda_{\mu}^{\hspace{1mm}\nu}[U(x)A_{\nu}(x)U^{-1}(x)-\frac{i}{g}\partial_{\nu}U(x)U^{-1}(x)],\\
\label{noether-lorentz-gauge-trans-s}
\tilde{\psi}(\tilde{x})&=&S[\Lambda]U(x)\psi(x),\hspace{2mm}\tilde{x}_{\mu}=\Lambda_{\mu}^{\hspace{1mm}\nu}x_{\nu}.
\end{eqnarray}
By Noether's theorem, we know that a symmetry of a Lagrangian yields a conserved current. Now we have a new kind of symmetry expressed by Eqs.~(\ref{noether-lorentz-gauge-trans-v}) and (\ref{noether-lorentz-gauge-trans-s}), so we can expect that this new symmetry could yield some new kinds of conserved current. Under an infinitesimal Lorentz transformation, we have
\begin{eqnarray}
\label{noether-lorentz-trans-small-x}
\tilde{x}_{\mu}=x_{\mu}+\delta{x}_{\mu}, \hspace{2mm}\delta{x}_{\mu}=\omega_{\mu\nu}x^{\nu},
\end{eqnarray}
where $\omega_{\mu\nu}=-\omega_{\nu\mu}$ are infinitesimal parameters. By Eqs.~(\ref{noether-lorentz-gauge-trans-v}) and (\ref{noether-lorentz-gauge-trans-s}), the fields transform as
\begin{eqnarray}
\label{noether-lorentz-gauge-trans-v-small}
\tilde{A}^{\mu}(x+\delta{x})&=&
[\delta_{\hspace{1mm}\nu}^{\mu}-\frac{i}{2}\omega_{\alpha\beta}(\mathcal{J}^{\alpha\beta})^{\mu}_{\hspace{1mm}\nu}]
[U(x)A^{\nu}(x)U^{-1}(x)-\frac{i}{g}\partial^{\nu}U(x)U^{-1}(x)],\\
\label{noether-lorentz-gauge-trans-s-small}
\tilde{\psi}(x+\delta{x})&=&
[1-\frac{i}{2}\omega_{\alpha\beta}\mathcal{S}^{\alpha\beta}]U(x)\psi(x),
\hspace{2mm}\widetilde{\bar{\psi}}(x+\delta{x})=
\bar{\psi}(x)U^{-1}(x)[1+\frac{i}{2}\omega_{\alpha\beta}\mathcal{S}^{\alpha\beta}],
\end{eqnarray}
where $(\mathcal{J}^{\alpha\beta})_{\mu\nu}=i(\delta^{\alpha}_{\hspace{1mm}\mu}\delta^{\beta}_{\hspace{1mm}\nu}
-\delta^{\alpha}_{\hspace{1mm}\nu}\delta^{\beta}_{\hspace{1mm}\mu})$ and $\mathcal{S}^{\alpha\beta}=\frac{i}{4}[\gamma^{\alpha},\gamma^{\beta}]$. For infinitesimal transformations, the unitary matrix $U(x)$ can be parameterized as
\begin{eqnarray}
\label{noether-lorentz-trans-small-u}
U(x)=\mathrm{exp}(igy_{\mu}N^{\mu}),  \hspace{2mm}y_{\mu}=\delta{x}_{\mu}=\omega_{\mu\nu}x^{\nu}.
\end{eqnarray}
Here $N^{\mu}(x)=N_{\mu}^{a}(x)T^{a}$ are functions taking values in the generators of the $SU(N)$ Lie algebra, whose meanings will be discussed later on. With this designation, Eq.~(\ref{noether-lorentz-gauge-trans-v-small}) can be expanded as
\begin{eqnarray}
\label{noether-lorentz-gauge-trans-v-small-expand}
\tilde{A}_{\mu}(x+\delta{x})&=&B_{\mu}(x)+\mathcal{O}(\omega^2),\\
B_{\mu}(x)&=&A_{\mu}(x)+\omega_{\mu\beta}(A^{\beta}(x)-N^{\beta}(x))
+y^{\beta}\{\partial_{\mu}N_{\beta}(x)-ig[A_{\mu}(x),N_{\beta}(x)]\}.\nonumber
\end{eqnarray}
We can define the variation $\Delta{A}_{\mu}(x)=\tilde{A}_{\mu}(x+\delta{x})-A_{\mu}(x)$ and the variation at a single point $\delta{A}_{\mu}(x)=\tilde{A}_{\mu}(x)-A_{\mu}(x)$, then these two kinds of variations are related by
\begin{eqnarray}
\label{noether-var-rela}
\Delta{A}_{\mu}(x)=\delta{A}_{\mu}(x)+\delta{x}^{\beta}\partial_{\beta}A_{\mu}(x).
\end{eqnarray}
Using this equation, we can derive the field variation at a single point as
\begin{eqnarray}
\label{noether-field-var-gauge}
\delta{A}_{\mu}(x)&=&\Delta{A}_{\mu}(x)-\delta{x}^{\beta}\partial_{\beta}A_{\mu}(x),\\
&=&-y^{\beta}\partial_{\beta}A_{\mu}(x)+\omega_{\mu\beta}(A^{\beta}(x)-N^{\beta}(x))\\
&+&y^{\beta}\{\partial_{\mu}N_{\beta}(x)-ig[A_{\mu}(x),N_{\beta}(x)]\}+\mathcal{O}(\omega^2),\nonumber
\end{eqnarray}
which can be easily recombined into a more compact formulation
\begin{eqnarray}
\label{noether-field-var-gauge-re}
\delta{A}_{\mu}(x)=y^{\beta}F_{\mu\beta}+\omega_{\mu\beta}(A^{\beta}-N^{\beta})
-y^{\beta}\{\partial_{\mu}(A_{\beta}-N_{\beta})-ig[A_{\mu},A_{\beta}-N_{\beta}]\}+\mathcal{O}(\omega^2).
\end{eqnarray}
Similarly, we derive the variation at a single point for fermion fields
\begin{eqnarray}
\label{noether-field-var-fermion}
\delta{\psi}(x)&=&\tilde{\psi}(x)-{\psi}(x)
=-\frac{i}{2}\omega_{\alpha\beta}\mathcal{S}^{\alpha\beta}{\psi}(x)-y^{\beta}[\partial_{\beta}-igN_{\beta}]\psi(x)+\mathcal{O}(\omega^2),\\
\label{noether-field-var-fermion-bar}
\delta{\bar{\psi}}(x)&=&\widetilde{\bar{\psi}}(x)-{\bar{\psi}}(x)
=\frac{i}{2}{\bar{\psi}}(x)\omega_{\alpha\beta}\mathcal{S}^{\alpha\beta}-y^{\beta}[\partial_{\beta}{\bar{\psi}}(x)+ig{\bar{\psi}}(x)N_{\beta}]
+\mathcal{O}(\omega^2).
\end{eqnarray}
For the infinitesimal transformations (\ref{noether-lorentz-trans-small-x})-(\ref{noether-lorentz-gauge-trans-s-small}) and the Lagrangian~(\ref{qcd-lag}), the Noether theorem asserts that
\begin{eqnarray}
\label{noether-field-var-eq}
0=\Delta&=&\partial_{\mu}(\delta{x}^{\mu}\mathcal {L})+\frac{\partial\mathcal {L}}{\partial{A_{\mu}}}\delta{A}_{\mu}+\frac{\partial\mathcal {L}}{\partial(\partial_{\mu}A_{\nu})}\delta(\partial_{\mu}A_{\nu})\nonumber\\
&+&\frac{\partial\mathcal {L}}{\partial{\psi}}\delta{\psi}+\delta\bar{\psi}\frac{\partial\mathcal {L}}{\partial\bar{\psi}}
+\frac{\partial\mathcal {L}}{\partial(\partial_{\mu}\psi)}\delta(\partial_{\mu}\psi)+\delta(\partial_{\mu}\bar{\psi})\frac{\partial\mathcal {L}}{\partial(\partial_{\mu}\bar{\psi})}.
\end{eqnarray}
Employing the properties of variation at a single point $\delta(\partial_{\mu}A_{\nu})=\partial_{\mu}\delta A_{\nu}$,  $\delta(\partial_{\mu}\psi)=\partial_{\mu}\delta\psi$ and $\delta(\partial_{\mu}\bar{\psi})=\partial_{\mu}\delta\bar{\psi}$, Eq.~(\ref{noether-field-var-eq}) can be reformulated as
\begin{eqnarray}
\label{noether-field-var-eq-re}
\Delta&=&\partial_{\mu}{\mathcal{J}}^{\mu}+\mathcal {E}=0,\\
\label{noether-field-var-eq-re-j}
{\mathcal{J}}^{\mu}&=&\delta{x}^{\mu}\mathcal {L}+\frac{\partial\mathcal {L}}{\partial(\partial_{\mu}A_{\nu})}\delta{A}_{\nu}+\frac{\partial\mathcal {L}}{\partial(\partial_{\mu}\psi)}\delta\psi+\delta\bar{\psi}\frac{\partial\mathcal {L}}{\partial(\partial_{\mu}\bar{\psi})},\\
\label{noether-field-var-eq-re-e}
\mathcal {E}&=&\left[\frac{\partial\mathcal {L}}{\partial{A_{\nu}}}-\partial_{\mu}\frac{\partial\mathcal {L}}{\partial(\partial_{\mu}A_{\nu})}\right]\delta{A}_{\nu}+\left[\frac{\partial\mathcal {L}}{\partial{\psi}}-\partial_{\mu}\frac{\partial\mathcal {L}}{\partial(\partial_{\mu}\psi)}\right]\delta\psi+\delta\bar{\psi}\left[\frac{\partial\mathcal {L}}{\partial\bar{\psi}}-\partial_{\mu}\frac{\partial\mathcal {L}}{\partial(\partial_{\mu}\bar{\psi})}\right].
\end{eqnarray}
By imposing Euler-Lagrange equations, we have $\mathcal {E}=0$, and then we obtain the conserved current
\begin{eqnarray}
\label{noether-field-var-eq-re-j-con}
\partial_{\mu}{\mathcal{J}}^{\mu}=0.
\end{eqnarray}
For the infinitesimal transformations (\ref{noether-lorentz-trans-small-x}), (\ref{noether-field-var-gauge-re})-(\ref{noether-field-var-fermion-bar}) and the Lagrangian~(\ref{qcd-lag}), the current ${\mathcal{J}}^{\mu}$ can be calculated to be
\begin{eqnarray}
\label{noether-field-var-eq-re-j-form}
{\mathcal{J}}^{\mu}&=&\frac{1}{2}\omega_{\alpha\beta}M^{\mu\alpha\beta},\\
\label{noether-field-var-eq-re-m-form}
M^{\mu\alpha\beta}&=&M_{qs}^{\mu\alpha\beta}+M_{qo}^{\mu\alpha\beta}+M_{gs}^{\mu\alpha\beta}+M_{go}^{\mu\alpha\beta}.
\end{eqnarray}
That is, the generators $M^{\mu\alpha\beta}$ can be divided into four parts naturally. The parts that can be identified to be quark spin and orbital angular momentums are given by
\begin{eqnarray}
\label{noether-quark-spin}
M_{qs}^{\mu\alpha\beta}&=&\frac{1}{2}\epsilon^{\mu\alpha\beta\rho}\bar{\psi}\gamma_{\rho}\gamma_{5}\psi,\\
\label{noether-quark-orbit}
M_{qo}^{\mu\alpha\beta}&=&\frac{i}{2}[\bar{\psi}\gamma^{\mu}x^{\alpha}(\partial^{\beta}-igN^{\beta})\psi-(\alpha\longleftrightarrow
\beta)]+\mathrm{H.C.}\\
&+&({\eta}^{\mu\alpha}x^{\beta}-{\eta}^{\mu\beta}x^{\alpha}){\mathcal{L}}_{\mathrm{quark}},\nonumber
\end{eqnarray}
while the parts that can be identified to be gluon polarization and orbital angular momentums are given by
\begin{eqnarray}
\label{noether-gluon-spin}
M_{gs}^{\mu\alpha\beta}&=&-2\mathrm{Tr}[F^{\mu\alpha}(A^{\beta}-N^{\beta})-F^{\mu\beta}(A^{\alpha}-N^{\alpha})],\\
\label{noether-gluon-orbit}
M_{go}^{\mu\alpha\beta}&=&2\mathrm{Tr}[F^{\mu}_{\hspace{2mm}\nu}(F^{\nu\alpha}x^{\beta}-F^{\nu\beta}x^{\alpha})]\\
&+&2\mathrm{Tr}[x^{\beta}F^{\mu}_{\hspace{2mm}\nu}D^{\nu}(A^{\alpha}-N^{\alpha})-(\alpha\longleftrightarrow
\beta)]\nonumber\\
&+&({\eta}^{\mu\alpha}x^{\beta}-{\eta}^{\mu\beta}x^{\alpha}){\mathcal{L}}_{\mathrm{gluon}}.\nonumber
\end{eqnarray}
Here $D_{\alpha}=\partial_{\alpha}-ig[A_{\alpha},\cdot~]$. ${\mathcal{L}}_{\mathrm{quark}}$ and ${\mathcal{L}}_{\mathrm{gluon}}$ are respectively the corresponding gauge invariant fermion parts and boson parts of the Lagrangian~(\ref{qcd-lag}). In above calculations, we have let the $\theta$ term to be zero, because it does not yield interesting results. The effects of the field $N^{\mu}$ can be revealed now. For $N^{\beta}=0$, we obtain the type of decomposition of Jaffe and Manohar~\cite{Jaffe:1989jz}. For $N^{\beta}=A^{\beta}$, we obtain the type of decomposition of Ji~\cite{Ji:1996ek}; A remarkable point is that the term~(\ref{noether-gluon-spin}), which describes the gluon polarization, vanishes in this case; This type of decomposition is gauge invariant. For $N^{\beta}=A^{\beta}_{\mathrm{pure}}$, we obtain a covariant version of the type of decompositions of Chen et al.~\cite{Chen:2008ag}.  Another point we should mention is that no surface terms are subtracted or added in the foregoing derivations. The expressions~(\ref{noether-quark-spin})-(\ref{noether-gluon-orbit}) are the straightforward results of Eqs.~(\ref{noether-field-var-gauge-re})-(\ref{noether-field-var-fermion-bar}) and (\ref{noether-field-var-eq-re-j}). Of course, a term $H^{\mu\alpha\beta}$ which satisfies $\partial_{\mu}H^{\mu\alpha\beta}=0$ can be added into the definition of $M^{\mu\alpha\beta}$, which does not spoil the conserved feature of the current ${\mathcal{J}}^{\mu}$~\cite{Belinfante:1940q}. This kind of surface term can lead to the type of decompositions of Wakamatsu~\cite{Wakamatsu:2010cb}. Even the gauge invariant extension of Chern-Simons current~(\ref{chern-simons-current-gie}) can emerge by adding appropriate surface terms.

The derivations of gauge invariant decompositions of the angular momentum from the Noether theorem are also discussed in~\cite{Zhou:2011z} and~\cite{Lorce:2013gxa} through different methods. We shall give another derivation from a different perspective in a subsequent paper~\cite{Guo:2013jia}.

\section{The Problem of Gauge Dependence}\label{sec:4}

We have defined the pure gauge field $A^{\mu}_{\mathrm{pure}}$ in section~\ref{sec:2}, which is used to constructed the gauge invariant extension of the Chern-Simons current in section~\ref{sec:2} and is also used to formulate gauge invariant descriptions of the gluon spin in section~\ref{sec:4}. Despite of the superficial success of the foregoing gauge invariant constructions, a question that should be mentioned is whether these constructions really yield gauge-independent results. We focus on the Chern-Simons current, which is tractable because of the involved totally anti-symmetrical tensor. This question can be more easily addressed in the (1+1)-dimensional~(2D) Schwinger model. The Lagrangian of 2D Schwinger model is
\begin{eqnarray}
\label{schw-lag}
\mathcal{L}=-\frac{1}{4}F_{\mu\nu}F^{\mu\nu}+i\bar{\psi}\gamma^{\mu}(\partial_{\mu}+ieA_{\mu})\psi,
\end{eqnarray}
which is a $U(1)$ gauge theory. Its Chern-Simons current is
\begin{eqnarray}
\label{schw-lag-cs}
K^{\mu}=\frac{1}{2}\varepsilon^{\mu\nu}A_{\nu},\hspace{2mm}\partial_{\mu}K^{\mu}=\frac{1}{4}\varepsilon^{\mu\nu}F_{\mu\nu}.
\end{eqnarray}
A gauge invariant extension of Eq.~(\ref{schw-lag-cs}) is
\begin{eqnarray}
\label{schw-lag-cs-gie}
\mathcal{K}^{\mu}=\frac{1}{2}\varepsilon^{\mu\nu}(A_{\nu}-\partial_{\nu}\theta)
=\frac{1}{2}\varepsilon^{\mu\nu}\left(A_{\nu}-\partial_{\nu}\frac{1}{\partial^2}\partial^{\alpha}A_{\alpha}\right).
\end{eqnarray}
Here we have employed Eq.~(\ref{pure-gauge-laudau-phys-solution}) to express the pure gauge field manifestly. The 2D Schwinger model can be solved  exactly by bosonization, that is, we can make the replacements
\begin{eqnarray}
\label{schw-lag-cs-boson}
i\bar{\psi}\gamma^{\mu}\partial_{\mu}\psi=\frac{1}{2}\partial_{\mu}\phi\partial^{\mu}\phi,\hspace{2mm}
\bar{\psi}\gamma^{\mu}\psi=\frac{1}{\sqrt{\pi}}\varepsilon^{\mu\nu}\partial_{\nu}\phi,\hspace{2mm}
\bar{\psi}\gamma^{\mu}\gamma^5\psi=\frac{1}{\sqrt{\pi}}\partial^{\mu}\phi.
\end{eqnarray}
The calculations involved in~(\ref{schw-lag-cs-gie}) can be most conveniently implemented in the Lorentz gauge. However, following Manohar~\cite{Manohar:1990eu}, we perform the calculations in the axial gauge, which here is a more persuasive argument for the gauge-independence of the current~(\ref{schw-lag-cs-gie}). In the axial gauge $n_{\mu}A^{\mu}=0$, and we can solve the gauge field as
\begin{eqnarray}
\label{schw-lag-axial-form}
A^{\mu}=\varepsilon^{\mu\nu}n_{\nu}\sigma.
\end{eqnarray}
By Eq.~(\ref{schw-lag-cs-boson}) and~(\ref{schw-lag-axial-form}), we obtain the reduced Lagrangian
\begin{eqnarray}
\label{schw-lag-axial}
\mathcal{L}=\frac{1}{2}\partial_{\mu}\phi\partial^{\mu}\phi+m\sigma{n}_{\mu}\partial^{\mu}\phi
+\frac{1}{2}\left[{n}_{\mu}\partial^{\mu}\sigma\right]^2.
\end{eqnarray}
Here $m=\frac{e}{\sqrt{\pi}}$ is the well known mass of the gauge field in the Schwinger model. The propagators of the Lagrangian~(\ref{schw-lag-axial}) can be derived by the functional method, which is done explicitly in~\cite{Manohar:1990eu}. For the gauge field~(\ref{schw-lag-axial-form}), the current $\mathcal{K}^{\mu}$ in Eq.~(\ref{schw-lag-axial-form}) is
\begin{eqnarray}
\label{schw-lag-cs-gie-axis}
\mathcal{K}^{\mu}=\frac{1}{2}\varepsilon^{\mu\nu}
\left(\varepsilon_{\nu\beta}{n}^{\beta}\sigma
-\partial_{\nu}\frac{1}{\partial^2}{n}^{\beta}\varepsilon_{\alpha\beta}\partial^{\alpha}\sigma\right).
\end{eqnarray}
With the propagators that are given in~\cite{Manohar:1990eu}, we obtain
\begin{eqnarray}
\label{schw-lag-cs-gie-axis-corr}
\langle{0\vert} T(\mathcal{K}^{\mu}\phi)\vert {0}\rangle&=&\frac{im}{2}\varepsilon^{\mu\nu}
\left(\varepsilon_{\nu\beta}{n}^{\beta}-\frac{1}{q^2}q_{\nu}{n}^{\beta}\varepsilon_{\alpha\beta}q^{\alpha}\right)
\frac{1}{(n^{\alpha}q_{\alpha})(q^2-m^2)}\nonumber\\
&=&\frac{1}{2}\frac{im}{(n^{\alpha}q_{\alpha})(q^2-m^2)}\frac{\varepsilon^{\mu\nu}q^{\alpha}}{q^2}(l_{\nu}q_{\alpha}-l_{\alpha}q_{\nu}),
\hspace{2mm}l_{\alpha}=\varepsilon_{\alpha\beta}{n}^{\beta},
\end{eqnarray}
which is gauge-dependent at first sight. But for any vectors $a_{\mu}$ and $b_{\nu}$ in 2D, we have
\begin{eqnarray}
\label{schw-lag-cs-gie-axis-vec}
a_{\mu}b_{\nu}-a_{\nu}b_{\mu}=(\varepsilon^{\alpha\beta}a_{\beta}b_{\alpha})\varepsilon_{\mu\nu}, \hspace{2mm}
\varepsilon^{\alpha\mu}\varepsilon_{\alpha\nu}=-\delta^{\mu}_{\nu},
\end{eqnarray}
which means that
\begin{eqnarray}
\label{schw-lag-cs-gie-axis-vec-1}
l_{\nu}q_{\alpha}-l_{\alpha}q_{\nu}=(\varepsilon^{\mu\beta}l_{\beta}q_{\mu})\varepsilon_{\nu\alpha}
=(\varepsilon^{\mu\beta}\varepsilon_{\beta\lambda}{n}^{\lambda}q_{\mu})\varepsilon_{\nu\alpha}
=(n^{\beta}q_{\beta})\varepsilon_{\nu\alpha}.
\end{eqnarray}
This leads to the gauge-independent result
\begin{eqnarray}
\label{schw-lag-cs-gie-axis-corr-resu}
\langle{0\vert} T(\mathcal{K}^{\mu}\phi)\vert {0}\rangle=\frac{1}{2}\frac{im}{q^2(q^2-m^2)}{q}^{\mu}.
\end{eqnarray}
This means that the matrix element
\begin{eqnarray}
\label{schw-lag-cs-gie-axis-corr-resu-me}
\langle{0\vert} \mathcal{K}^{\mu}\vert {\phi(q)}\rangle=\lim_{\substack{q^2\rightarrow {m^2}}}(q^2-m^2)\langle{0\vert} T(\mathcal{K}^{\mu}\phi)\vert {0}\rangle=\frac{i}{2m}{q}^{\mu},
\end{eqnarray}
which is well defined and gauge-independent.

The above results of the Schwinger model suggest that similar calculations in four dimensional QCD may be also well defined and gauge-independent. We try to calculate the forward matrix element of the current $\mathcal{K}^{\mu}$ in Eq.~(\ref {chern-simons-current-gie}), in which the pure gauge field is given by Eq.~(\ref{pure-gauge-laudau-phys-solution})
\begin{eqnarray}
\label{pure-gauge-laudau-phys-solution-pure}
A^{\mu}_{\mathrm{pure}}=\partial^{\mu}\frac{1}{\partial^2}\partial^{\alpha}A_{\alpha}+\mathcal{O}(g),
\end{eqnarray}
and the forward matrix element is taken between massive quark states $\langle{p,s\vert} \mathcal{K}^{\mu}\vert {p,s}\rangle$.
For four dimensional QCD, we do the calculations in the Lorentz gauge.  Obviously, the one-loop calculation is independent of the gauge parameter $\xi$, because the physical field $A^{\mu}_{\mathrm{phys}}=A^{\mu}-A^{\mu}_{\mathrm{pure}}=A^{\mu}-\partial^{\mu}\frac{1}{\partial^2}\partial^{\alpha}A_{\alpha}+\mathcal{O}(g)$ is a transverse field. The calculation with the dimensional regularization yields
\begin{eqnarray}
\label{chern-simons-forward-me}
\langle{p,s\vert} \mathcal{K}^{\mu}\vert {p,s}\rangle=\frac{\alpha_s}{2\pi}\left(\frac{7}{2}C_{F}+\frac{3}{2}C_{F}N_{\epsilon}\right)\bar{u}(p,s)\gamma^{\mu}\gamma^{5}u(p,s),
\end{eqnarray}
where $\alpha_s=\frac{g^2}{4\pi}$, $C_{F}=\frac{N_{c}^{2}-1}{2N_{c}}$, $d=4-2\epsilon$, $N_{\epsilon}=\frac{1}{\epsilon}-\gamma_{E}+\mathrm{Log}4\pi+\mathrm{Log}\frac{\mu^2}{M^2}$ and $\mu$ is the renormalization scale. We have used the on-shell relation $\gamma^{\mu}{p}_{\mu}u(p)=Mu(p)$ and $p^2=M^2$.

\section{Conclusions}\label{sec:5}

In this paper, we proposed a possible candidate of gauge invariant description of the gluon spin. We constructed a gauge invariant extension of the Chern-Simons current in section~\ref{sec:2}, and we verified its gauge-independence in section~\ref{sec:4}. This extended Chern-Simons current has several remarkable features: It is gauge invariant and is also Lorentz covariant; It satisfies the same equation as the conventional Chern-Simons Current; Furthermore, we shall show that it can describe the spin of the Laguerre-Gauss Laser mode in a subsequent paper~\cite{Guo:2013jia}. These features support the proposal that it can be considered as an appropriate description of the gluon spin.

We also discussed the relations among three decompositions of the nucleon spin through the Noether theorem in section~\ref{sec:3}. The canonical angular momentum current can be derived from the Noether theorem. However, for gauge theories, the canonical conserved current is not gauge invariant. This problem can be cured by considering the Noether current induced by a generalized Lorentz transformation. The origin of this generalized lorentz transformation could be illuminated as follows. From the conventional lore, only the space-time components of the fields are rotated under Lorentz transformations. But when we consider fields with internal local freedom, under the Lorentz transformation, the internal local freedom and the space-time components of the fields can all be rotated without breaking the symmetry of the system. This leads to a generalized Lorentz symmetry, which yields a new kind of Noether current. This Noether current has the nice property that the induced decompositions can accommodate different decompositions of the nucleon spin in a single framework. This generalized symmetry introduce a new field $N^{\mu}$, whose effects are quite similar to the role of the Stueckelberg field~\cite{Stueckelberg:1938zz}, in the meaning of that it is regarded as a compensating field to formulate gauge invariant expressions. Recently, by analyzing the Wigner distributions, Burkardt~\cite{Burkardt:2012sd} proposed that the difference between the gauge invariant extension~\cite{Hatta:2011zs,Hatta:2011ku} of Jaffe-Manohar's quark orbital angular momentum and that of Ji can be understood as a torque acting on a quark. Tentatively, we can speculate that this torque may be related to the field $N^{\mu}$. Its effects could be uncovered along a similar way.

\acknowledgments

We thank P. M. Zhang, D. G. Pak and M. Siddikov for useful discussions. This work was supported in part by Fondecyt~(Chile) grant 1100287 and by Project Basal under Contract No.~FB0821.

\bibliographystyle{utphys}

\bibliography{SpinGluonRef}

\end{document}